\documentclass[%
 reprint,%
 amssymb, amsmath,%
 prc,cha,%
superscriptaddress
]{revtex4-2}
\usepackage[colorlinks=true,linkcolor=blue]{hyperref}%
\usepackage{physics}
\usepackage{graphicx}
\usepackage{lipsum}
\usepackage{mwe}
\usepackage{caption}
\usepackage{dcolumn}% Align table columns on decimal point
\usepackage{bm}% bold math
\usepackage{adjustbox}
\usepackage{rotating}
\usepackage{xcolor}
\usepackage{ulem}
\begin{document}
\preprint{APS/123-QED}
\title{Electric and magnetic $\gamma$-ray strength functions at finite-temperature} 

\author{Amandeep Kaur}\email[]{akaur.phy@pmf.hr}
\affiliation{Department of Physics, Faculty of Science, University of Zagreb, Bijeni\v{c}ka c. 32,  10000 Zagreb, Croatia}

\author{Esra Y{\"{u}}ksel}\email[]{e.yuksel@surrey.ac.uk}
\affiliation{School of Mathematics and Physics, University of Surrey, Guildford, Surrey GU2 7XH, United Kingdom}

\author{Nils Paar}\email[]{npaar@phy.hr}
\affiliation{Department of Physics, Faculty of Science, University of Zagreb, Bijeni\v{c}ka c. 32,  10000 Zagreb, Croatia}

\date{\today}

\begin{abstract} 
The $\gamma$-ray strength function ($\gamma$SF) is essential for understanding the electromagnetic response in atomic nuclei and modeling astrophysical neutron capture rates. We introduced a microscopic description of both electric dipole (E1) and magnetic dipole (M1) $\gamma$SFs that includes finite-temperature effects within relativistic density functional theory. The temperature dependence of the total  electromagnetic $\gamma$SFs shows significant modification in the low-energy region due to thermal unblocking effects, essential for agreement with recent particle-$\gamma$ coincidence data from the Oslo method.  An investigation of the electric and magnetic contributions to the total $\gamma$SF in hot nuclei indicates that the M1 mode becomes more prominent in the low-energy region, different than what is known at zero temperature. This microscopic approach offers new insights into the interplay between E1 and M1 $\gamma$SFs at finite-temperature, and opens new perspectives for future studies of $(n,\gamma)$ reactions and nucleosynthesis in hot stellar environments.

\end{abstract}
\maketitle
%\tableofcontents
%\section{\label{sec:level1}First-level heading:\protect\\ The line
%break was forced \lowercase{via} \textbackslash\textbackslash}

%%%%%%%%%%%%%%%%%%%%%%%%%%%%%%%%%%%%%%%%%%%%%%%%%%%%%%%%%%%%%%%%%%%%%%%%%%
%\section{Introduction} \label{Intro}
%%%%%%%%%%%%%%%%%%%%%%%%%%%%%%%%%%%%%%%%%%%%%%%%%%%%%%%%%%%%%%%%%%%%%%%%%
%$\textit{Introduction.}$
Gamma-ray strength functions ($\gamma$SFs), which characterize the excitation and deexcitation of atomic nuclei through $\gamma$-ray absorption and emission, are crucial for modeling astrophysical phenomena \cite{CAPOTE20093107,KLanganke_2001}. These functions, particularly those including electric dipole (E1) and magnetic dipole (M1) strengths, serve as critical inputs for reaction rate calculations in the statistical Hauser-Feshbach model, shaping our understanding of nucleosynthesis and the origin of nuclear elements in the universe \cite{PhysRev.87.366,10.1063/1.1945212,koning2023talys}. According to the Brink-Axel hypothesis, the $\gamma$SF is independent of nuclear excitation energy, remaining unchanged for both cold nuclei in the ground state and hot nuclei in thermally excited states \cite{brink,PhysRev.126.671}. However, the experimentally observed low-energy enhancement, also called `upbend' \cite{goriely2019reference,PhysRevC.71.044307, PhysRevC.106.034314,PhysRevC.82.014318,PhysRevC.109.054311,PhysRevC.81.024319,PhysRevC.87.014319}, deviates from this prediction. Although the Brink-Axel hypothesis may hold for giant resonances, it appears to be invalid for the lowest energy transitions, where the low-energy strength is redistributed with changes in the nucleus's excitation energy \cite{sieja2023brink}. Therefore, the temperature ($T$) of the hot nucleus must be considered to gain deeper insight into the nature of $\gamma$SFs. Phenomenological models often incorporate a $T$-dependent width to estimate the low-energy strength. However, this approach lacks a strong theoretical foundation, as the low-energy strength arises from underlying physical mechanisms \cite{CAPOTE20093107}. 

A microscopic description of the finite-temperature effects using the (quasiparticle) random phase approximation ((Q)RPA) can serve as a valuable tool for probing transition probabilities between individual excited states in both the initial and final nuclei. Several versions of temperature-dependent (Q)RPA are available to analyze separately E1 \cite{NIU2009315,PhysRevC.96.024303} and M1 \cite{PhysRevC.109.024305,PhysRevC.110.014307} transitions. However, no comprehensive study within the (Q)RPA framework has simultaneously addressed the temperature effects on both E1 and M1 components of the $\gamma$SFs. Various shell model studies have also explored low-energy enhancement in M1 strength, primarily in medium-mass nuclei \cite{PhysRevLett.111.232504,PhysRevC.98.064312,PhysRevC.110.054313}. Within the shell model, for the first time, the low-energy $\gamma$SF has been studied for $^{44}$Sc by incorporating both electric and magnetic dipole contributions \cite{PhysRevLett.119.052502}, showing that the low-energy enhancement in $\gamma$SF arises from M1 transitions. However, distinct experimental data for E1 and M1 components of the $\gamma$SF at low energies are limited, making it difficult to compare with model calculations directly. 

In this Letter, we introduce a novel microscopic method to describe $\gamma$SFs, including both E1 and M1 transitions across a broad mass range of nuclei, with explicitly included finite-temperature effects within the framework of the relativistic energy density functional (REDF). In this way, we study for the first time the temperature evolution of both electric and magnetic $\gamma$SFs and discuss the results in view of recent particle-$\gamma$ coincidence data from the Oslo method \cite{PhysRevC.109.054311,goriely2019reference,PhysRevLett.127.182501,PhysRevC.96.024313}. Theoretical framework is based on the fully self-consistent finite-temperature relativistic quasiparticle RPA (FT-RQRPA) developed within the REDF \cite{PhysRevC.109.014314,PhysRevC.109.024305}.
The framework also incorporates deexcitations from highly excited states of a hot nucleus, using the principle of detailed balance \cite{RING1984261,PhysRevC.100.024307}. We investigate the individual contributions of the E1 and M1 components to the total $\gamma$SF, particularly in the low-energy region. This allows us, for the first time, to study the temperature evolution of both electric and magnetic $\gamma$SFs and to discuss the results in light of recent particle-$\gamma$ coincidence data obtained using the Oslo method \cite{PhysRevC.109.054311,goriely2019reference,PhysRevLett.127.182501,PhysRevC.96.024313}. Compared to previous approaches in the EDF framework that do not include M1 transitions, or to the other approaches that incorporate $T$-effects at a phenomenological level, this work provides a consistent description of $\gamma$SFs for both E1 and M1 transitions. The REDF employed  in this study explicitly includes density-dependence in the interaction vertex functions, which is essential to provide a reasonable description of the symmetry energy that governs the properties of E1 excitation spectra and the corresponding $\gamma$SFs. Whereas the non-linear meson-exchange interactions cannot accurately describe the $\gamma$SFs due to overestimated symmetry energy, which leads to an overprediction of the E1 transition strength \cite{PhysRevC.109.054311}.

In this work, the properties of even–even closed- and open-shell spherical nuclei are described within the finite-temperature Hartree–Bardeen–Cooper–Schrieffer (FT-HBCS) framework \cite{GOODMAN198130, yuksel2014effect}, and excited states are obtained using the FT-RQRPA. In the calculations, we employ the relativistic density-dependent point-coupling interaction DD-PCX, which is particularly well-suited for modeling E1 transitions due to its constraints from dipole polarizability data \cite{PhysRevC.99.034318}.
The point-coupling REDF is derived from the Lagrangian density, $\mathcal{L} = \mathcal{L}_{\textrm{PC}} + \mathcal{L}_{\textrm{IV-PV}}$. Here, $\mathcal{L}_{\textrm{PC}}$ includes fermion contact interaction terms in the isoscalar-scalar, isoscalar-vector, and isovector-vector channels \cite{PhysRevC.99.034318,NIKSIC20141808}. The Lagrangian density $\mathcal{L}$ also includes the relativistic isovector–pseudovector (IV-PV) contact interaction, which contributes to the FT-RQRPA residual interaction for unnatural-parity M1-type excitations \cite{PhysRevC.102.044315,kruvzic2023magnetic}.
A separable form of the pairing interaction is employed in both the FT-HBCS and FT-RQRPA frameworks \cite{PhysRevC.80.024313}. 

The occupation probabilities of single-particle states in a hot nucleus are given by $n_i=v_{i}^{2}(1-f_{i})+u_{i}^{2}f_{i}$, where $u_i$ and $v_i$ are the BCS amplitudes. The $T$-dependent Fermi-Dirac distribution function is defined as $f_{i}=[1+exp(E_{i}/k_{\textrm{B}}T)]^{-1}$, where $k_{\textrm{B}}$ is the Boltzmann constant, and $E_{i}$ denotes the quasiparticle energy of state. The non-charge exchange FT-RQRPA matrix is given by \cite{PhysRevC.109.014314,PhysRevC.109.024305},
\begin{equation}
\left( { \begin{array}{cccc}\label{eq:qrpa}
 \widetilde{C} & \widetilde{a} & \widetilde{b} & \widetilde{D} \\
 \widetilde{a}^{+} & \widetilde{A} & \widetilde{B} & \widetilde{b}^{T} \\
-\widetilde{b}^{+} & -\widetilde{B}^{\ast} & -\widetilde{A}^{\ast}& -\widetilde{a}^{T}\\
-\widetilde{D}^{\ast} & -\widetilde{b}^{\ast} & -\widetilde{a}^{\ast} & -\widetilde{C}^{\ast}
 \end{array} } \right)
 \left( {\begin{array}{cc}
\widetilde{P}  \\
\widetilde{X }  \\
\widetilde{Y}  \\
\widetilde{Q} 
 \end{array} } \right)
 = E_{w}
  \left( {\begin{array}{cc}
\widetilde{P}  \\
\widetilde{X}  \\
\widetilde{Y}  \\
\widetilde{Q} 
\end{array} } \right), \end{equation}
where $ E_{w}$ denotes the excitation energies, and $\widetilde{P}, \widetilde{X}, \widetilde{Y}, \widetilde{Q}$ represent the corresponding eigenvectors. Finally, temperature-dependent reduced transition probability for T$J=$ E1 and M1 transitions is calculated as $B(\textrm{T}J,w)=\bigl|\langle w ||\hat{F}_{J}||\widetilde0\rangle \bigr|^{2}$,
\begin{widetext}
\begin{equation}
\begin{split}
B(\textrm{T}J,w)=\bigl|\langle w ||\hat{F}_{J}||\widetilde0\rangle \bigr|^{2}
&=\biggl|\sum_{c\geq d}\Big\{(\widetilde{X}_{cd}^{w} + (-1)^{j_{c}-j_{d}+J}\widetilde{Y}_{cd}^{w})(u_{c}v_{d}+(-1)^{J}v_{c}u_{d})\sqrt{1-f_{c}-f_{d}} \\
&+(\widetilde{P}_{cd}^{w}+(-1)^{j_{c}-j_{d}+J}\widetilde{Q}_{cd}^{w})(u_{c}u_{d}-(-1)^{J}v_{c}v_{d})\sqrt{f_{d}-f_{c}}\Big\}\langle c ||\hat{F}_{J}||d\rangle\biggr|^{2},
\end{split}
\label{bel}
\end{equation}
\end{widetext}
where $|w\rangle$ is the excited state and $|\widetilde0\rangle$ is the correlated FT-RQRPA vacuum state, and $\hat{F}_{J}$ is the transition operator of the relevant excitation. 
%More details are given in Refs. \cite{PhysRevC.109.014314,PhysRevC.109.024305}.

Within the FT-RQRPA, the excitation spectra of nuclei are represented as discrete sets of peaks having two-quasiparticle structure, and additional couplings with more complex configurations are required to describe the spreading widths, such as in the second RPA (SRPA) or particle-vibration coupling models. However, the inclusion of finite-temperature effects in these approaches, using density-dependent REDFs that are necessary for the study of $\gamma$SF, still remains challenging. The self-energy insertions on particle and hole states spread the resonances and shift their centroids \cite{Daoutidis2012}. Following  microscopic study within SRPA theory including also $2p2h$ excitations \cite{Drozdz1990,Daoutidis2012}, the spreading effects in E1 transition strength can be approximated by folding the FT-RQRPA strength distribution $R_{E1}(E)=\sum L(E,w) B(E1)$ with a Lorentzian function,
\begin{equation}\label{L1}
       L(E,w)=\frac{1}{2\pi}\frac{\Gamma(E)}{(E-w-\Delta(E))^2+\Gamma(E)^2/4},
\end{equation}
where $w$ is the excitation energy of the FT-RQRPA response, and the $\Delta(E)$ and $\Gamma(E)$ are the real and complex part of the self-energy as $\Sigma (E)=\Delta (E)+i\Gamma(E)/2$. $\Gamma(E)$ denotes energy-dependent width, that can be obtained from measured decay width of particle $\gamma_p$ and hole $\gamma_h$ states \cite{Drozdz1990}, and is given by $\Gamma(E)=\frac{1}{E}\int_0^E d\epsilon[\gamma_p(\epsilon)+\gamma_h(\epsilon-E)](1+C_{G}).$
Using the collective width $\Gamma(E)$, the energy shift $\Delta(E)$ of the self energy can be obtained using the dispersion relation as 
  $\Delta(E)=\frac{1}{2\pi} \mathcal{P} \int^{\infty}_{-\infty}dE' \frac{\Gamma(E')}{E'-E}(1+C_{E})$ \cite{Drozdz1990},
assuming different interference coefficients $C_G$ and $C_E$ in the real and complex parts of the self-energy. The $\gamma$SF is fine tuned by adjusting these interference coefficients. In this way, the spreading effects are included empirically, avoiding explicit calculation of couplings to complex configurations at the SRPA level that are beyond reach for systematic calculations required for astrophysical applications \cite{Daoutidis2012}.
E1 photoabsorption (excitation) $\gamma$SF ($f_{E1}$) ($MeV^{-3}$) is deduced from the following relation
\begin{equation}\label{fE1}
f_{E1}(E)=\frac{16\pi e^2}{27\hbar^3c^3}\times R_{E1}(E).
\end{equation}
Since there is very limited data on M1 photoabsorption $\gamma$SFs ($f_{M1}$), it is not feasible to use the aforementioned procedures of interference coefficients for the M1 case. Even the available $f_{M1}$ data from the inelastic proton scattering ($p,p'$) or other experiments are highly fragmented, and the Lorentzian functions merely reproduce the overall shapes and total integrated M1 strengths \cite{goriely2019reference}. However, M1 contribution must be taken into account for a complete description of the low-energy region of $\gamma$SFs. Thus, we used a basic Lorenztion function (at constant width $\Gamma=$ 1 MeV) of spin-flip resonance for the consistent modelling of M1 photoabsorption $\gamma$SFs  ($f_{M1}) (MeV^{-3})$,  
\begin{equation}\label{fM1}
f_{M1}(E)=\frac{16\pi}{27\hbar^3c^3} 
\frac{\Gamma/2\pi}{((E-w)^2+\Gamma^2/4)}B(M1),
\end{equation}
where $B(M1)$ is the magnetic dipole strength distribution given in the units of $\mu_N^2 MeV^{-1}$. Further, by applying the principle of detailed balance, the photoemission (deexcitation) $\gamma$-ray strength function $\overleftarrow{f}$($MeV^{-3}$), for electromagnetic transitions at finite temperature can be derived from the photoabsorption $\gamma$SFs as \cite{PhysRevC.100.024307}
\begin{equation}\label{deexc_function}
\overleftarrow{f}_{E1,M1}(E,T)=\frac{1}{1-exp(-E/T)}\times f_{E1,M1}(E).
\end{equation}
This additional factor, applied to the photoabsorption $\gamma$SFs, which modifies the low-energy region at finite temperatures and make the strength finite at energies approaching $E_{\gamma}\rightarrow$ 0 MeV.
%%%
\begin{figure}[htp]
\includegraphics[width=\linewidth,clip=true]{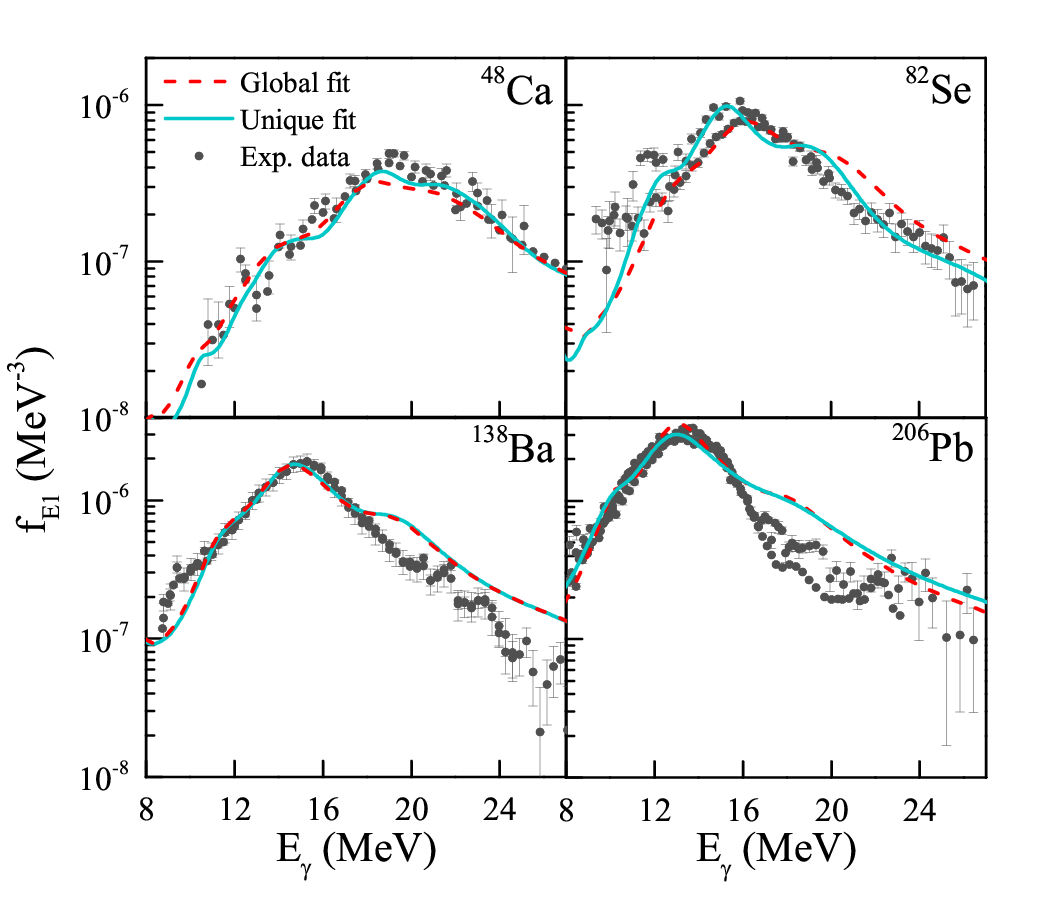}
   \vspace{-0.8cm}
  \caption{The E1 photoabsorption $\gamma$SFs ($f_{E1}$) for several medium- and heavy-mass nuclei at $T = 0$ MeV, obtained using unique and global fits, compared with experimental data \cite{goriely2019reference}.}
  \label{fig1} 
\end{figure}
%%%
%%%
\begin{figure*}[htp]
\includegraphics[width=\linewidth,clip=true]{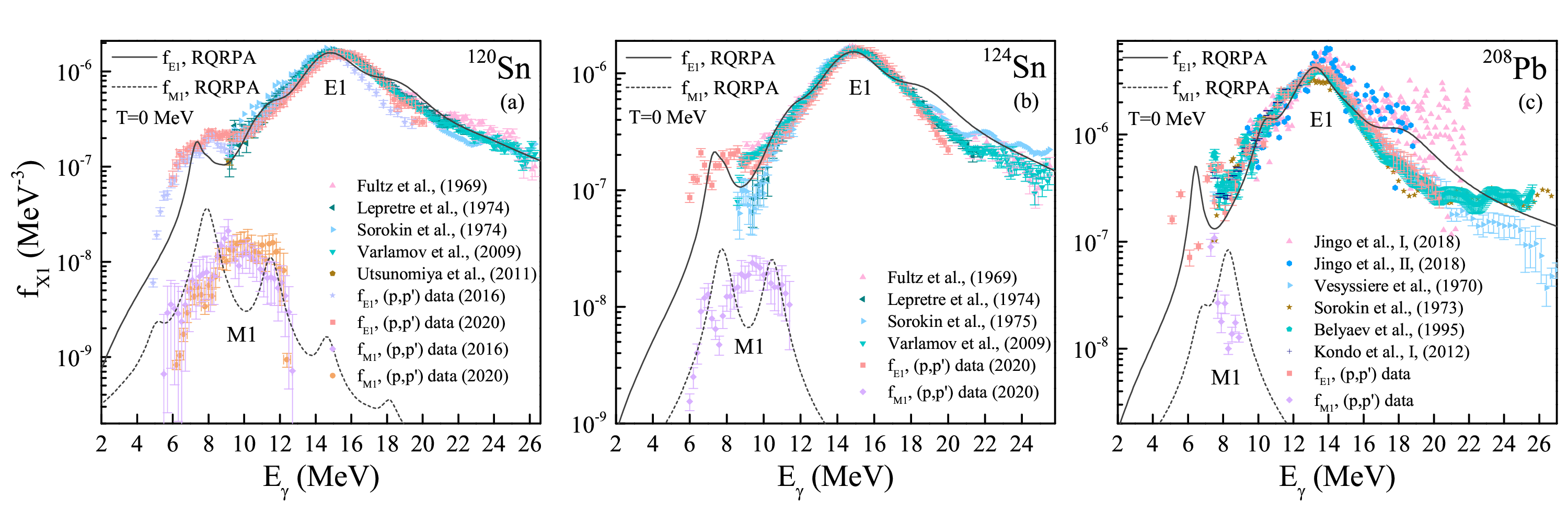}
   \vspace{-0.8cm}
  \caption{The $\gamma$SFs for E1 and M1 transitions in (a) $^{120}$Sn, (b) $^{124}$Sn, and (c) $^{208}$Pb nuclei calculated within the RQRPA framework at $T=$ 0 MeV, and compared with the experimental photoabsorption/photoneutron $\gamma$SFs by Fultz \textit{et al.} \cite{PhysRev.186.1255}, Lepretre \textit{et al.} \cite{LEPRETRE197439}, Sorokin \textit{et al.} \cite{Sorokin}, Varlamov \textit{et al.} \cite{varlamov2010evaluated}, Utsunomiya \textit{et al.} \cite{PhysRevC.84.055805}, Jingo \textit{et al.} I and II \cite{jingo2018studies}, Vesyssiere \textit{et al.} \cite{VEYSSIERE1970561}, Belyaev \textit{et al.} \cite{Belyaev}, Kondo \textit{et al.} \cite{PhysRevC.86.014316}; and ($p,p'$) data from Refs. \cite{goriely2019reference,KRUMBHOLZ20157,PhysRevC.102.034327}.}
  \label{fig2} 
\end{figure*}
%%%

%%%
To benchmark the FT-RQRPA description of $\gamma$SFs for E1 transitions ($f_{E1}$), we first performed calculations in the zero temperature limit for a set of 42 spherical or nearly spherical nuclei covering a broad range of mass number, $i.e.$, $A=$ 40$-$208. We constrained the interference coefficients $C_G$ and $C_E$ associated with the resonance width and energy shift of the E1 strength, respectively, by minimizing the $\chi^2$ objective function to reproduce the experimental photoabsorption/photoneutron $\gamma$SF. 
The experimental data, compared with all the FT-RQRPA calculations in this work, is taken from the recent database for $\gamma$SFs \cite{goriely2019reference}. 
Two methods are used to constrain the interference coefficients $C_G$ and $C_E$, to introduce the broadening effects in the E1 transition strength. Method I, referred as global fit, computes the constant values of $C_G=$ -0.584 and $C_E=$ -1.149, for all nuclei considered to achieve the best fit to the experimental data. In method II, called unique fit, individual unique values of interference coefficients for each nucleus are constrained by the experimental photoabsorption/photoneutron $\gamma$SF, see table \cite{figshare}.
To compare the two methods and calculate the widths, along with the corresponding experimental data \cite{goriely2019reference}, Fig. \ref{fig1} shows the RQRPA results for the E1 photoabsorption $\gamma$SFs ($f_{E1}$) using the global and unique methods to constrain the widths for cold nuclei ($T = 0$ MeV). The photoabsorption $f_{E1}$ in the high-energy or the GDR region, calculated using Eq. (\ref{fE1}), are shown for medium to heavier masses. The figure demonstrates that, while there are minor differences between the global and unique methods, both methods show strong agreement with the experimental data. For further analysis of the $f_{E1}$, individual interference coefficients for each nucleus are extracted from the unique fitting procedure.
\begin{figure}[ht]
    %\centering
    \resizebox{!}{0.58\textwidth}{
    \includegraphics{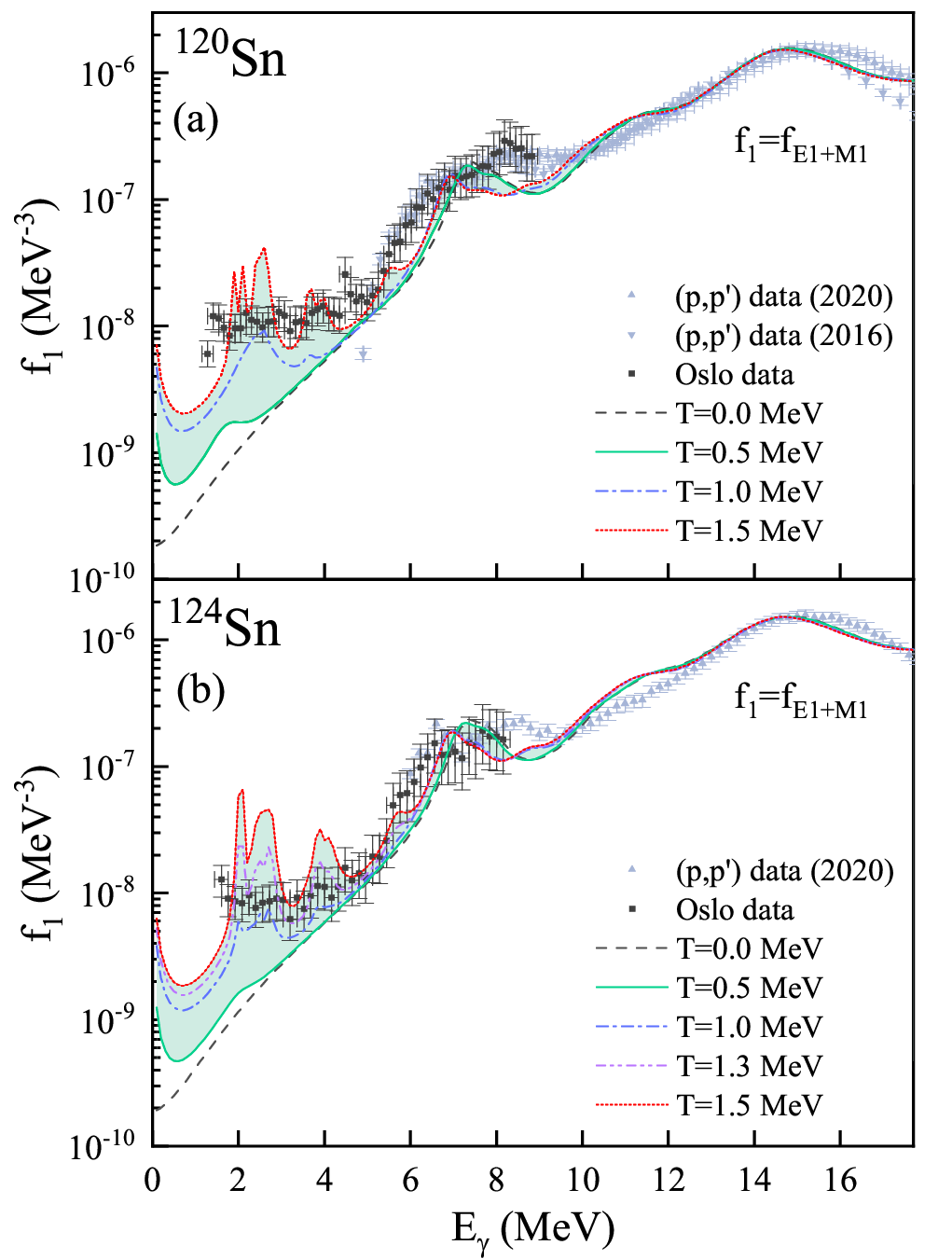}} \vspace{-0.1cm}
    \caption{Temperature evolution of FT-RQRPA calculated E1+M1 $\gamma$SFs ($f_1$) from $T=0$ to 1.5 MeV is shown for (a) $^{120}$Sn and (b) $^{124}$Sn, with comparisons to experimental data from Oslo \cite{goriely2019reference,PhysRevLett.127.182501} and ($p,p'$) measurements \cite{KRUMBHOLZ20157,PhysRevC.102.034327}.}
    \label{fig3}
\end{figure}
\begin{figure*}[ht]
    %\centering
    \resizebox{!}{0.32\textwidth}{
    \includegraphics{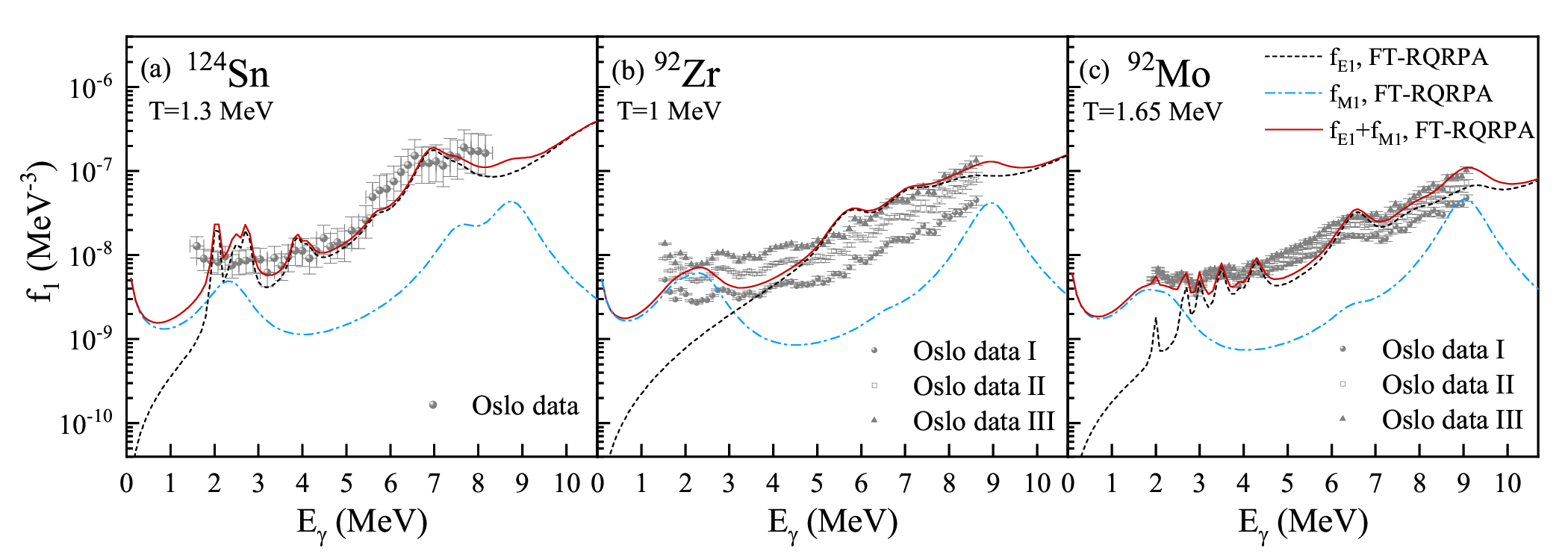}} \vspace{-0.3cm}
    \caption{Individual FT-RQRPA contributions of E1 and M1 components in total $\gamma$-ray strength functions at low-energies for (a) $^{124}$Sn, (b) $^{92}$Zr and (c) $^{92}$Mo at finite-temperatures, and compared with Oslo data sets \cite{goriely2019reference,PhysRevC.96.024313}.}
    \label{fig4}
\end{figure*}
%%%

In Fig. \ref{fig2}, the $f_{E1}$ and $f_{M1}$ photoabsorption $\gamma$SFs at $T=$ 0 MeV are illustrated for $^{120}$Sn, $^{124}$Sn and $^{208}$Pb nuclei, along with the corresponding experimental photoabsorption/photoneutron and ($p,p'$) data. The figure shows that the RQRPA calculated $f_{E1}$ in the high-energy region (i.e., the GDR region) exhibits good comparison with experimental $\gamma$SFs. In the low-energy region, the RQRPA calculations show a peak-like structure due to pygmy dipole strength, which is also observed in the ($p,p'$) data at energies $E_{\gamma} \approx 6 - 10$ MeV for the nuclei under consideration. As discussed earlier, experimental data on $f_{M1}$ is very limited and fragmented. Therefore, a basic Lorentzian function with a constant width $\Gamma=$ 1 MeV is applied for the consistent modeling of $f_{M1}$ using Eq. (\ref{fM1}). The calculated $f_{M1}$ is also plotted in Fig. \ref{fig2} for $^{120}$Sn, $^{124}$Sn, and $^{208}$Pb, for which experimental M1 data from ($p,p'$) experiments are available. For the Sn isotopes, the experimental M1 data show two distinct peaks and similar features are reproduced in the RQRPA calculations. However, for $^{208}$Pb, the experimental M1 data consists of only a few scattered points, all of which lie under the calculated $f_{M1}$ curve. These results suggest that the Lorentzian parametrizations considered for the E1 and M1 $\gamma$SFs provide a good description of the experimental data in the ground state.

To further investigate the effect of temperature on the total $\gamma$SFs ($f_1 = f_{E1} + f_{M1}$), the deexcitation  $\gamma$SFs are calculated for the $^{120}$Sn and $^{124}$Sn nuclei at temperatures ranging from $T = 0$ to 1.5 MeV, as shown in Figs. \ref{fig3} (a) and (b). The results of FT-RQRPA are compared with the total $\gamma$SF data from ($p,p'$) experiments \cite{KRUMBHOLZ20157,PhysRevC.102.034327} and particle-$\gamma$ coincidence data from the Oslo method (also called Oslo data) \cite{goriely2019reference, PhysRevLett.127.182501}. We begin the discussion with the $T = 0$ MeV case, where the high-energy GDR region shows excellent agreement with the ($p,p'$) data for both Sn nuclei. The Oslo data is also consistent with the low-energy region, particularly the $E_{\gamma} \approx 4$–10 MeV range, which is important for astrophysical applications. It should be worth mentioning here that Oslo data, measured for $E_{\gamma}$ range between $S_n$ and $S_n-$ 2 or 3 MeV, includes total dipole $\gamma$SF, and it does not distinguish between E1 and M1 components. The M1 spin-flip resonance thus requires additional experimental constraints, which are highly desired. However, the decomposition of E1 and M1 components is available for $^{120}$Sn and $^{124}$Sn nuclei from inelastic proton scattering experiments, i.e.  ($p,p'$) reactions as the comparison shown in Figs. \ref{fig2} (a) and (b). It can be seen from these figures that the pygmy dipole strength dominates the tail of the gamma strength function, especially above 4 MeV. However, no enhancement `or' upbend is seen as we approach $E_{\gamma}=$ 0 MeV at zero temperature; instead, the $f_1$ curve gets smaller (see Fig. \ref{fig3}). 

After gaining insight into the dynamics of total $\gamma$SFs at $T=$ 0 MeV, we next analyze the evolution of $\gamma$SFs at finite temperatures. The results of Oslo experiments focus on deexcitation strength functions, whose theoretical explanation remains a topic of active research. Therefore, it is interesting to compare low-energy $\gamma$SFs calculated using FT-RQRPA at various temperatures with the Oslo data. Furthermore, contrary to the Brink hypothesis, several studies also demonstrate that the deexcitation strength function may differ from the photoabsorption strength function, which can be explained by a $T$-dependent correction to the photoabsorption strength function \cite{CAPOTE20093107,goriely2019reference}. However, the current work incorporates $T$-effects at the microscopic level to better describe $\gamma$SFs than phenomenological approaches. The finite temperature dependent deexcitation $\gamma$SF can be obtained from Eq. (\ref{deexc_function}). To evaluate the influence of finite temperature on $\gamma$SFs, the $f_{1}$ is shown in Fig. \ref{fig3} with increasing temperature from $T=$ 0.5 to 1.5 MeV. It is observed that the high-energy region of $f_1$ shows negligible change for the considered temperature range. However, the low-energy region is highly sensitive to temperature effects. As expected, the deexcitation $\gamma$SFs at finite temperatures in the low-energy region deviates significantly from the photoabsorption strength at zero temperature, particularly below $E_{\gamma}\approx$ 5 MeV. When $E_{\gamma}$ approaches towards 0 MeV, a non-zero behavior `or' an upbend is observed. The $T$-dependent factor $1/[1-exp(-E/T)]$ applied on both E1 and M1 strength has effectively enhanced the low-energy part of the strength, and this factor has negligible impact on the high-energy region. 

Various studies have been conducted to understand the nature of this upbend behavior \cite{PhysRevLett.111.232504,PhysRevLett.93.142504,PhysRevLett.119.052502}. It is expected that the very low-lying energy region of deexcitation $\gamma$SF is dominated by M1 transitions. Furthermore, a recent experimental study on pygmy dipole strength in hot nuclei highlights the significant influence of temperature on this phenomenon \cite{wieland2024extra}. Fig. \ref{fig4} provides a more detailed analysis of the nature of the upbend and the roles that E1 and M1 transitions play in the low-energy enhancement. The $f_{E1}$ and $f_{M1}$ components of the total $\gamma$SFs are distinguished for the $^{124}$Sn, $^{92}$Zr, and $^{92}$Mo nuclei at temperatures $T = 1.3$, 1.0, and 1.65 MeV, respectively. 
For $^{92}$Zr and $^{92}$Mo nuclei, Oslo data I, II and III represent the lower, recommended and upper limits of experimental measurements normalized within or at the limits of the range of uncertainties \cite{goriely2019reference,PhysRevC.96.024313}.
It is observed that the total FT-RQRPA $\gamma$SFs show a good agreement with the Oslo data. At finite-temperatures, new excitation channels open due to thermally occupied states above the Fermi level, which modify the electromagnetic response of nuclei, especially in the low-energy region \cite{PhysRevC.109.014314,PhysRevC.109.024305}. Consequently, the $E_{\gamma}=$ 0$-$3 MeV part of the $\gamma$SF shows the interplay between E1 and M1 components. This lower-energy region can be accounted for by taking only the M1 contribution, while the E1 component adds a small value to the $\gamma$SF at low energy but does not produce any additional enhancement. Although there is currently no experimental confirmation of this, it aligns with predictions from shell model studies \cite{PhysRevLett.111.232504, PhysRevLett.119.052502, PhysRevC.98.064312}.
To conclude, the microscopic analysis of $\gamma$SFs, as shown in Figs. \ref{fig3} and \ref{fig4}, indicate that temperature effects can significantly impact the electromagnetic response of nuclei in the low-energy region. This influence is particularly pronounced for M1 transitions at very low energies, which are essential for understanding the behavior of low-energy $\gamma$SFs.

%%%
In conclusion, the microscopic FT-RQRPA calculations of electromagnetic $\gamma$SFs introduced in this work resolve many limitations from previous studies, which often neglect magnetic components, lack a microscopic foundation of the finite-temperature effects, overestimate nuclear symmetry energy resulting in overvalued E1 response, or have restricted nuclear mass range.
The evolution of E1 and M1 $\gamma$SFs at finite-temperatures illustrates a low-energy enhancement of $\gamma$SF with increasing temperature due to thermal unblocking of new excitation channels. The analysis of individual contributions of electromagnetic components in total $\gamma$SF indicates that M1 strength accounts for the low-energy enhancement in $\gamma$SF. 
The interplay between E1 and M1 modes at finite-temperatures is crucial for understanding and describing the experimental $\gamma$SF in the low-energy region, and thus it could substantially impact (n,$\gamma$) cross-sections and neutron capture rates, making it a compelling topic for the future large-scale calculations of complete $\gamma$SFs and nuclear reaction studies for astrophysical applications. Additionally, further experimental studies are required to provide separate insight into the E1 and M1 components in $\gamma$SFs.

%%%
We thank S. Goriely and S. K{\"{u}}\c{c}{\"{u}}ksucu for stimulating discussion and for helping to clarify important aspects in modeling gamma strength functions. This work is supported by the Croatian Science Foundation under the project Relativistic Nuclear Many-Body Theory in the Multimessenger Observation Era (HRZZ-IP-2022-10-7773). E.Y. acknowledges support from the UK STFC under award no. ST/Y000358/1.

\bibliographystyle{apsrev4-2}
\bibliography{GSF}
\end{document}